\documentclass{appolb}
\usepackage{epsfig}
\begin{document}

\title{First Results on $K^+$ Production in $pp$ and $p$D Interactions from  
ANKE and
\\Planned Experiments on the Light Scalar Resonances $a_0/f_0$(980) 
at COSY\thanks{Work supported by: BMBF-WTZ, DFG, Polish State Committee 
for Scientific Research, Russian Academy of Science, Russian Academy
of Science.  
  }%
}

\author{M.~B\"uscher\\(representing the ANKE collaboration\thanks{For a complete collaboration list see 
{\tt www.fz-juelich.de/ikp/anke}})
\address{Institut f\"ur Kernphysik, 
         Forschungszentrum J\"ulich, 52425 J\"ulich} }
\maketitle

\begin{abstract}
ANKE is a magnetic spectrometer and detection system at an internal
target position of COSY-J\"ulich optimized for charged kaon detection.
Recent results from ANKE on kaon production in $pp$ and $p$D
interactions are reported. From the $pp$ data first absolutely
normalized angular and invariant-mass spectra for the reaction $pp\to
dK^+\bar K^0$ have been obtained.  A partial-wave decomposition
reveals a strong contribution of $S$-wave $K\bar K$-pairs with low
relative energy, suggesting dominance of resonant kaon production via
the $a_0^+$(980).  This indicates that systematic studies of the light
scalar resonances $a_0/f_0$(980) are possible at COSY. Final goal of
these measurements --- requiring a neutral-particle detector which is
not yet available --- is to obtain information about the
charge-symmetry breaking $a_0$-$f_0$ mixing.  From the analysis of the
$p$D data it is concluded that the $K^+$-production cross section on
the neutron is significantly larger as compared to the proton. A
cross-section ratio of $\sigma_n /
\sigma_p\sim 4$ is deduced.
\end{abstract}

\section{The ANKE spectrometer}
The COoler SYnchrotron COSY-J\"ulich~\cite{cosy}, which provides
proton beams in the energy range $T_p = 0.04 - 2.83$~GeV, is well
suited for the study of $K^+$-meson production in $pp$ and $pA$
reactions. In measurements with thin and windowless internal targets,
secondary processes of the produced mesons can be neglected and,
simultaneously, sufficiently high luminosities are obtained. For the
measurements described here, a cluster-jet target
\cite{clustertarget} with hydrogen or deuterium as target material has
been used, providing areal densities of up to $\sim 5\times
10^{14}$~cm$^{-2}$. With proton beam intensities of a few $10^{10}$
luminosities of $\mathcal{L} > 10^{31}\,\mathrm{cm^{-2}s^{-1}}$ have
been achieved.

The ANKE spectrometer~\cite{ANKE_NIM,K_NIM} consists of three dipole
magnets, which separate forward-emitted charged reaction products from
the circulating proton beam and allow to determine their emission
angles and momenta.  $K^+$-mesons in the momentum range $p_K\sim 150 -
600$ MeV/c can be detected, the angular acceptance is $\pm 12^{\circ}$
horizontally and up to $\pm 7^{\circ}$ vertically.

Subthreshold $K^+$-production in $pA$ reactions has been the prime
motivation for building ANKE and the detection system for
$K^+$-mesons. This is a very demanding task because of the small
$K^+$-production cross sections, e.g.\ 39~nb for $p$C collisions at
1.0~GeV~\cite{pnpi}. The results of these measurements have been
published in Refs.\
\cite{1.0GeV_PRL,MB_Ksyst,MB_MESON2002,Potentials_PLB,K_corr_EPJA}.
In subsequent experiments ANKE has been used to study kaon production
in more elementary (i.e.\ $pp$ and $p$D) reactions as well. 

From the $pp$ data information about the production of the scalar
resonance $a_0^+$(980) close to the $K\bar K$ threshold has been
extracted, see Sect.\ \ref{sec:a0f0_anke}. This experiment can also be
regarded as a successful feasibility test for a longer experimental
program which has the final goal to determine the charge-symmetry
breaking $a_0$-$f_0$ mixing amplitude. These measurements are
motivated in Sects.\ \ref{sec:a0f0_motivation} and
\ref{sec:a0f0_future} and will require the use of a photon detector
which is not available at COSY yet. In Sect.\ \ref{sec:pn_anke} first
data from ANKE on $K^+$-production in $p$D interactions are
presented. These data show that deuterium can be used as an effective
neutron target for meson-production studies like, e.g., for some of
the planned measurements on $a_0/f_0$-production. The data also yield
novel information about the $K^+$-production cross section in $pn$
interactions.

\section{Investigation of $a_0/f_0$-resonance production at COSY}
\label{sec:a0f0}
\subsection{Physics case}
\label{sec:a0f0_motivation}
One of the primary goals of hadronic physics is the understanding of
the internal structure of mesons and baryons, their production and
decays, in terms of quarks and gluons. The non-perturbative character
of the underlying theory --- Quantum Chromo Dynamics (QCD) --- hinders
straight forward calculations. QCD can be treated explicitly in the
low momentum-transfer regime using lattice techniques \cite{lattice},
which are, however, not yet in the position to make quantitative
statements about the light scalars. Alternatively, QCD inspired
models, which use effective degrees of freedom, are to be used. The
constituent quark model is one of the most successful in this respect
(see e.g.\ \cite{quarkmodel}). This approach treats the lightest
scalar resonances $a_0/f_0$(980) as conventional $q\bar{q}$
states. However, they have also been identified with $K\bar{K}$
molecules \cite{KK_molecules} or compact $qq$-$\bar{q}\bar{q}$ states
\cite{4q_states}. It has even been suggested that at masses below 1.0
GeV a complete nonet of 4-quark states might exist
\cite{4q_nonet}. 

The existing data base is insufficient to conclude on the structure of
the light scalar mesons and additional observables are urgently called
for. In this context the charge-symmetry breaking (CSB) $a_0$-$f_0$
mixing plays an exceptional role since it is sensitive to the overlap
of the two wave functions. It should be stressed that, although
predicted to be large long ago \cite{achasov}, this mixing has not
unambiguously been identified yet in corresponding experiments.

\subsection{Measurement of the strange decay channels with ANKE}
\label{sec:a0f0_anke}
\begin{figure}[ht]
  \centering
  \resizebox{10cm}{9.2cm}{\includegraphics[scale=1]{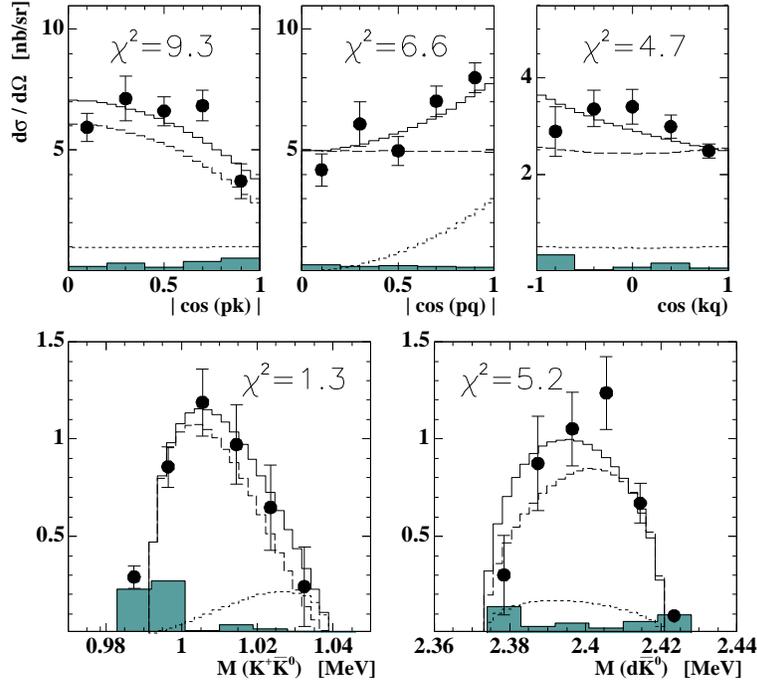}}
  \caption{ANKE data for the reaction $p(2.65\, \mathrm{GeV})p\to
  dK^+\bar{K}^0$ \cite{a+_PRL}. The shaded areas correspond to the
  systematic uncertainties of the acceptance correction. The dashed
  (dotted) line corresponds to $K^+\bar{K}^0$-production in a relative
  $S$- ($P$-) wave and the solid line is the sum of both
  contributions. For a definition of the angles $pk$, $pq$ and $kq$
  see Fig.\ \ref{fig:a+_vectors}.}  \label{fig:pp2dKKbar}
\end{figure}

An experimental program has been started at COSY which aims at
exclusive data on $a_0/f_0$ production close to the $K\bar{K}$
threshold from $pp$ \cite{cosy-11,a+_proposal}, $pn$, $pd$
\cite{momo,a0f0_proposal} and $dd$ \cite{css2002,dd_proposal}
interactions --- i.e.\ different isospin combinations in the initial
state. During the first experiment which has been made in this context
at ANKE, the reaction $pp {\to} dK^+\bar{K^0}$ has been measured
exclusively at beam energies of $T=2.65$ and 2.83 GeV, corresponding
to excess energies $Q=46$ and 106 MeV above the $K\bar K$
threshold. These measurements crucially depend on the high
luminosities achievable with internal targets, the large acceptance of
ANKE for close-to-threshold reactions, and the excellent kaon
identification with the ANKE detectors. The obtained differential
spectra for the lower beam energy are shown in Fig.\
\ref{fig:pp2dKKbar} \cite{a+_PRL}.

\begin{figure}[ht]
  \centering
  \includegraphics[scale=.5]{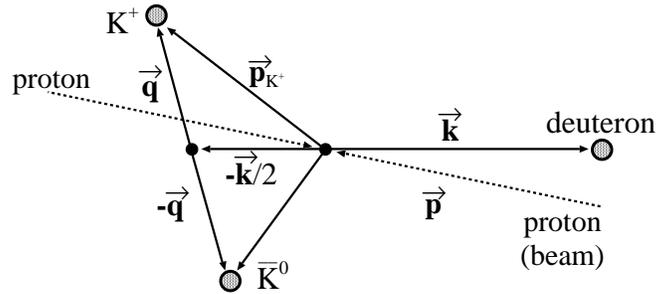}
  \caption{Definition of the vectors $\vec{p}$, $\vec{k}$ and
      $\vec{q}$ in the cms of the reaction $pp {\to}
      dK^+\bar{K^0}$. Angular distributions with respect to the beam
      direction $\vec{p}$ have to be symmetric around $90^\circ$ since
      the two protons in the entrance channel are indistinguishable.}
  \label{fig:a+_vectors}
\end{figure}

The background of misidentified events in the spectra of Fig.\
\ref{fig:pp2dKKbar} is less than 10\% which is crucial for the 
partial-wave analysis. This analysis reveals that the $K^+\bar{K}^0$
pairs are mainly (83\%) produced in a relative $S$-wave (dashed line
in Fig.\ \ref{fig:pp2dKKbar}), which has been interpreted in terms of
dominant $a_0^+$-resonance production, corresponding to a total cross
section of $\sigma(pp{\to}da_0^+ {\to}dK^+\bar{K}^0) = 83\%\cdot
\sigma(pp{\to}dK^+\bar{K}^0)= 32$~nb \cite{a+_PRL}. Based on these
data, which are in line with model predictions for different initial
isospin configurations \cite{brat}, it is concluded that the
production cross section for the light scalar resonances in hadronic
interactions is sufficiently large to permit systematic studies at
COSY (during our first beam time $\sim$1000 events have been collected
within five days of beam time using a hydrogen target with an average
luminosity of $L= 2.7\cdot 10^{31}\, \mathrm{cm}^{-2}
\mathrm{s}^{-1}$).  

The data from the second measurement at $Q=106$ MeV are still being
analyzed. As the next step of the experimental program a measurement
of the reaction $pn\to dK^+K^-$ at $Q\sim100$ MeV will be performed in
Feb.\ 2004 \cite{a0f0_proposal}. For these measurements deuterium will
be used as target material serving as an effective neutron target
The results of a similar experiment on the reaction $pn{\to}K^+X$ ---
demonstrating the feasibility of such experiments --- are described in
Sect.\ \ref{sec:pn_anke}. According to our cross-section estimates a
measurement of the reaction $dd\to\alpha K^+K^-$ should be feasible
within few weeks of beam time and is foreseen for winter 2004/05
\cite{css2002,dd_proposal}.

\subsection{Outline of future experiments using a photon detector}
\label{sec:a0f0_future}
Both, the $a_0^0$- and the $f_0$-resonances can decay into $K^+K^-$
and $K_SK_S$, whereas in the non-strange sector the decays are into
different final states according to their isospin, $a_0^\pm\to
\pi^\pm\eta$, $a_0^0\to \pi^0\eta$ and $f_0\to \pi^0\pi^0$ or 
$\pi^+\pi^-$.  Thus, only the non-strange decay channels have defined
isospin and allow to directly discriminate the two mesons. It is also
only by measuring the non-strange decay channels that CSB can be
investigated.  As described in the following, these measurements
require the use of a photon detector for active $\pi^0$- or
$\eta$-meson identification.  With such a detector the strange decay
channels $a_0/f_0\to K_SK_S$ should be measured in parallel and the
results can be compared with those from ANKE for the charged kaons.

\begin{figure}[ht]
  \begin{center}
    \resizebox{10cm}{!}{\includegraphics[scale=1]{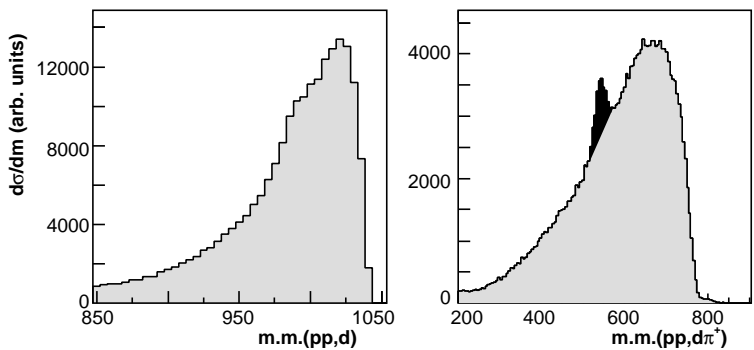}}
    \caption{ANKE data for the reaction $p(2.65\, \mathrm{GeV})p\to
    d\pi^+X$ \cite{a+_pieta}. Left: missing mass $m(pp,d)$ which
    contains the Flatt\'{e} distribution of the $a_0^+$ at a
    mass of $\sim980$ MeV/c$^2$; right: the missing mass
    $m(pp,d\pi^+)$ reveals the $\eta$ signal on top of a huge
    multi-pion background.}
  \label{fig:pp2dpieta}
 \end{center}
\end{figure}

Figure \ref{fig:pp2dpieta} shows the results from ANKE for the
reaction $p(2.65\, \mathrm{GeV}) p\to d \pi^+X$. The measurements have
been made in parallel to the ones for the decay channel $a_0^+\to
K^+\bar{K}^0$.  In contrast to these data, where the spectra contain
less than 10\% of misidentified particles, the $pp\to d\pi^+\eta$
signal is on top of a huge broad background stemming from multi-pion
events (right spectrum in Fig.\ \ref{fig:pp2dpieta}). This makes the
analysis of this channel much more demanding and even model dependent
\cite{a+_pieta}. A total cross section of  $\sigma(pp\to d\pi^+\eta)\sim 
4.6\, \mu$b  has been extracted from the data with
a resonant contribution of $\sigma(pp\to da_0^+\to d\pi^+\eta)\sim
1.1\, \mu$b.

\begin{figure}[ht]
  \begin{center} 
  \resizebox{7cm}{5.5cm}{\includegraphics[scale=1]{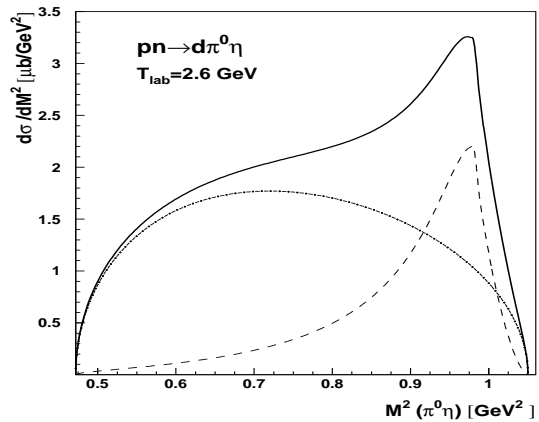}}
  \caption{Predicted invariant $\pi^0\eta$-mass spectrum
  \cite{a0_f0-mixing_PLB} for the reaction $pn\to d\pi^0\eta$ taking
  into account resonant production via the $a_0^0$ (dashed line) as
  well as non-resonant production (dashed dotted). The shape of the
  $a_0$-resonance has been described by a Flatt\'{e} distribution
  with: $K$-matrix pole at 999~MeV, $\Gamma_{a_0\to \pi\eta}=70$ MeV,
  $\Gamma_{KK}/\Gamma_{\pi\eta}=0.23$.}
  \label{fig:m_pi0eta_model}
 \end{center}
\end{figure}

The data from ANKE indicate that with better background suppression
(i.e.\ identification of the $\eta$ in the final state) the
$a_0$-resonance can be studied at COSY in the non-strange decay
channels as well.  Thus, for the proposed measurements on the
$a_0/f_0$ the detection of the photons from $\pi^0$ and $\eta$ decays
is required. Due to the larger $Q$ values in the non-strange channels
the angular acceptance of the corresponding photon detector should be
as large as possible. Figure \ref{fig:m_pi0eta_model} shows the
predicted invariant $\pi^0\eta$ mass spectrum for the reaction $pn\to
da_0^0$ \cite{a0_f0-mixing_PLB} assuming an ``ideal'' experiment
(i.e.\ perfect $\pi^0$ and $\eta$ identification and no background ---
comparable to the current conditions at ANKE for $K^+$-mesons, c.f.\
lower left spectrum in Fig.\ \ref{fig:pp2dKKbar}). The
$a_0^0$-resonance is, in fact, visible on a broad background of
non-resonant $\pi^0\eta$ events.  Note that the calculated total cross
sections for the resonant and non-resonant contributions from Ref.\
\cite{a0_f0-mixing_PLB} are in accord with the above mentioned
experimental values from ANKE.

Since it is possible to manipulate the initial isospin of purely
hadronic reactions one can identify observables that vanish in the
absence of CSB \cite{miller,ANKE_WS}.  The idea behind the proposed
experiments is the same as behind recent measurements of CSB effects
in the reactions $np\to d\pi^0$ \cite{opper} and $dd\to\alpha\pi^0$
\cite{stephenson}. However, the interpretation of the signal from the 
scalar mesons is largely simplified as compared to the pion
case. Since the $a_0$ and the $f_0$ are rather narrow overlapping
resonances, the $a_0$-$f_0$ mixing in the final state is enhanced by
more than an order of magnitude compared to CSB in the production
operator (i.e.\ ``direct'' CSB violating $dd\to \alpha a_0$
production) and should, e.g., give the dominant contribution to the
CSB effect via the reaction chain $dd\to
\alpha f_0(I{=}0) \to \alpha a_0^0(I{=}1) \to \alpha (\pi^0\eta)$ 
\cite{CH}. This reaction seems to be most promising for the extraction 
of CSB effects, since the initial deuterons and the $\alpha$ particle
in the final state have isospin $I{=}0$ (``isospin filter''). Thus,
any observation of $\pi^0\eta$ production in this particular channel
is a direct indication of CSB and can give information about the
$a_0$-$f_0$ mixing amplitude \cite{CH}. According to our cross section
estimates, it should be possible to collect sufficient statistics
within a few weeks of beam time if a frozen-pellet target is used
offering luminosities of more than $10^{32}\, \mathrm{cm}^{-2}
\mathrm{s}^{-1}$ \cite{css2002,ANKE_WS}.

In analogy with the measurement of CSB effects in the reaction $np\to
d\pi^0$, it has been predicted that the measurement of angular
asymmetries (i.e.\ forward-backward asymmetry in the $da_0$ c.m.s.) 
can give information about the $a_0$-$f_0$ mixing
\cite{a0_f0-mixing_PLB,tarasov,kudr}. It was stressed in Ref.\ 
\cite{tarasov} that --- in contrast to the $np\to d\pi^0$ experiment 
where the forward-backward asymmetry was found to be as small as 0.17\%
\cite{opper} --- the reaction $pn\to d\pi^0\eta$ is subject to a
kinematical enhancement.  As a consequence, the effect is predicted to
be significantly larger in the $a_0$/$f_0$ case. The numbers range
from some 10\% \cite{tarasov} to factors of a few
\cite{a0_f0-mixing_PLB} and, thus, should easily be observable in an
experiment with a large acceptance photon detector at COSY. It has
been pointed out in Ref.\ \cite{kudr} that the analyzing power of the
reaction $\vec p n\to d \pi^0 \eta$ also carries information about the
$a_0$-$f_0$ mixing amplitude. This quantity can be measured at COSY as
well, using the polarized proton beam and a azimuthally symmetric
photon detector.

\section{$K^+$-meson production on neutrons}
\subsection{Physics case}
\label{sec:pn_motivation}
Experimental data on the $K^+$-production cross section from $pn$
interactions in the close-to-threshold regime are not available
yet. This quantity is, for example, crucial for the theoretical
description of $pA$ and $AA$ data since it has to be used as an input
parameter for corresponding model calculations, like transport
codes. Predictions for the ratio $\sigma_{n} / \sigma_{p}$ range from
one to six, depending on the underlying model assumptions: in Ref.\
\cite{Piroue} it has been proposed that there is no difference between
$K^+$ production on the neutron and proton, whereas the analysis in
Ref.\ \cite{Tsushima} yields $\sigma_{n} / \sigma_{p}\sim 2$ for the
total production cross sections. The authors of Ref.\ \cite{Wilkin}
draw an analogy between $K^+$- and $\eta$-meson production and give a
ratio of six for the ratio between production on the neutron and
proton.

\subsection{First results from ANKE}
\label{sec:pn_anke}
\begin{figure}[ht]
  \centering
  \resizebox{7cm}{5.5cm}{\includegraphics[scale=1.]{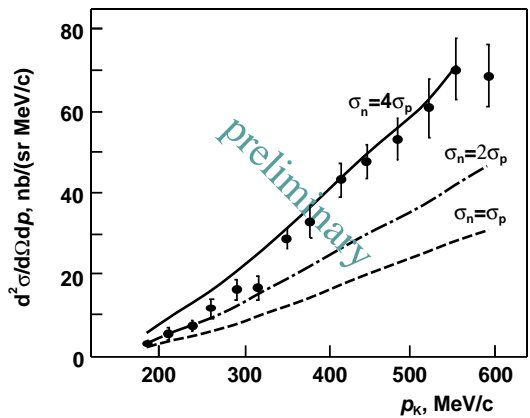}}
  \caption{Double differential $p\mathrm{D}\to K^+X$ cross section at
   2.02 GeV in comparison with our model calculations
  using different ratios $\sigma_n/\sigma_p$ (lines). The vertical and
  horizontal kaon emission angles have been restricted to
  $\vartheta<4^{\circ}$ during the analysis. The overall systematic 
  uncertainty from the luminosity normalization of 20\% is not included 
  in the error bars.} 
  \label{fig:pd2K+X}
\end{figure}

$K^+$-production in $p$D interactions has been investigated with ANKE
at two beam energies, $T_p=1.83$ and 2.02 GeV.  Figure
\ref{fig:pd2K+X} shows the $K^+$-momentum spectrum for the
higher beam energy. Based on the assumption that the $K^+$-production
cross section is governed by the sum of the elementary $pp$ and the
$pn$ cross sections, the spectra have been analyzed in a simple
phase-space approach, assuming $\sigma_{\mathrm D} = \sigma_{p} +
\sigma_{n}$ with $\sigma_n/\sigma_p$ being a free parameter. 
The main results of this analysis are described below, however, for
further details we refer to a forthcoming publication.

%According to the analysis in Ref.\ \cite{Tsushima}, six main reaction
%channels lead to kaon production on the deuteron at our beam
%energies. They are listed in Table \ref{tab:sibcr} together with the
%corresponding cross sections which have been calculated using the
%parameterizations given in Ref.\ \cite{Tsushima}.   

In order to determine $\sigma_{n} / \sigma_{p}$, phase-space
distributed $pp\to K^+X$ and $pn\to K^+X$ events have been generated
with the PLUTO package \cite{pluto} taking into account the intrinsic
motion of the nucleons in the deuteron.  The events have been
generated for all reaction channels which may lead to $K^+$-production
in $pN$ interactions at our beam energy and have been weighted
according to the cross-section parameterizations from Ref.\
\cite{Tsushima}. Each event subsequently has been tracked through the
spectrometer and all detection efficiencies have been taken into
account.  In Fig.\ \ref{fig:pd2K+X} we show the resulting momentum
spectra based on the approaches from from Ref.\ \cite{Piroue} (dashed
line labeled by ``$\sigma_n {=} \sigma_p$'') and Ref.\ \cite{Tsushima}
(dash-dotted line labeled by ``$\sigma_n {=} 2\sigma_p$'').

The apparent difference between the calculated and measured cross
sections can be due to the fact that the ratio $\sigma_n/\sigma_p$
is different than in Refs.\ \cite{Piroue,Tsushima}.  Thus we repeated
the simulations keeping the relative weights of the individual $pp$
and $pn$ channels constant (as given by Ref.\
\cite{Tsushima}) but treating the ratio of the sum of these two
contributions, i.e.\ $\sigma_{n} / \sigma_{p}$, as a free parameter.
The best agreement between data and calculations is obtained for
$\sigma_{n} / \sigma_{p} \sim 3$ at 1.83 GeV and $\sigma_{n} /
\sigma_{p} \sim 4$ at 2.02 GeV (solid line in Fig.\ \ref{fig:pd2K+X}).

\begin{figure}[ht]
  \centering 
  \resizebox{\textwidth}{3.5cm}{\includegraphics[scale=1.]{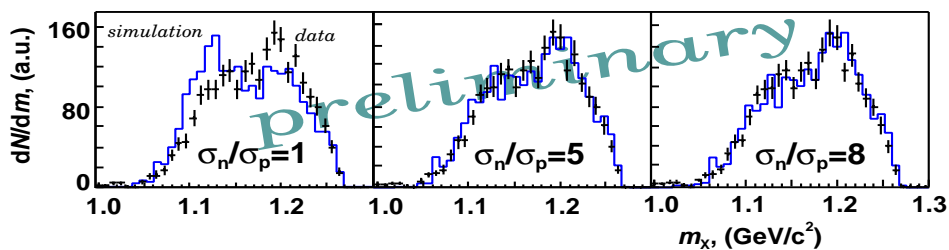}}
  \caption{Missing mass $m_X$ for
  $p\mathrm{D}\to K^+pX$ events at $T=
  2.02$ GeV in comparison with our model calculations using different
  ratios $\sigma_n/\sigma_p$ (lines).}  \label{fig:pd2K+pX}
\end{figure}

The resulting large cross-section ratio $\sigma_{n} / \sigma_{p}$ from
the inclusive spectra is supported by the analysis of missing-mass
spectra from $p\mathrm{D}\to K^+pX$ events recorded during the same
beam time. The spectrum for $T=2.02$ GeV is shown in Fig.\
\ref{fig:pd2K+pX} and is compared with the result of the Monte-Carlo
simulations, again for different ratios $\sigma_{n} / \sigma_{p}$. In
the simulations it has been taken into account that protons can either
stem from the $K^+$ production processes (e.g.\ $pp\to pK^+\Lambda$
but not from $pn\to nK^+\Lambda$) or from the subsequent hyperon decay
($pp$ and $pn$). The best agreement between data and simulations is
obtained for $\sigma_{n} / \sigma_{p} \sim (4-5)$.

\section{Acknowledgments}
The author is grateful to C.~Hanhart for contributing to parts of
Sect.\ \ref{sec:a0f0}, to Yu.~Valdau and V.~Koptev for supplying the
data on $p$D interactions, and to V.~Kleber and W.v.~Oers for
carefully reading the manuscript.

\end{document}